\documentclass[journal]{vgtc}                     

\onlineid{0}

\vgtccategory{Research}

\vgtcpapertype{please specify}

\title{Examining and Reflecting on A Community-Based Visualization Co-design Practice through Entanglement Theory}
\title{Interrogating and Refiguring A Community-based Visualization Co-design Practice with Entanglements}
\title{A Wake-up Call for Rethinking and (some of the time) Changing Epistemology in Visualization Research}
\title{From Epistemic Tensions to New Research Outcomes: Applying Entanglement Theory to Visualization Research}
\title{From Epistemic Tensions to Research Contributions: Reframing A Visualization Co-Design through Entanglement Theory}
\title{Epistemic Tensions: Reframing A Visualization Co-Design through Entanglement Theory}

\author{%
  Wei Wei,
  Foroozan Daneshzand, 
  Zezhong Wang,
  Erica Mattson,
  Jenny Farkas, 
  Sarah Storteboom,\\
  Charles Perin, and
  Sheelagh Carpendale
}

\authorfooter{
    \item
    Wei Wei and Charles Perin are with University of Victoria.
    E-mail: \{weiwei, cperin\}@uvic.ca.
    
    \item 
    Foroozan Daneshzand is with University of Calgary and Simon Fraser University. 
    E-mail: foroozan.daneshzand@ucalgary.ca
  
   \item
    Zezhong Wang is with University of St Andrews and Simon Fraser University.
    E-mail: wangzezhong2016@gmail.com.

    \item 
    Jenny Farkas is an artist, maker, writer, and creative economic developer.
  	E-mail: jfarkas@telus.net.

    \item 
    Sarah Storteboom is a data visualization specialist.
  	E-mail: sstorteboom@gmail.com.
    
    \item 
    Erica Mattson is an artist and a strategic advisor to artists.
  	E-mail: erica.mattson@gmail.com.

    \item 
    Sheelagh Carpendale is with Simon Fraser University.
    E-mail: sheelagh@sfu.ca.
}

\abstract{%
In this work, we present how employing the lens of entanglement helped us examine and reframe epistemic tensions arising in a visualization co-design project.
Entanglement theory challenges traditional assumptions in the visualization research community by emphasizing that knowledge is \emph{not} produced through linear, isolated processes, but is inherently entangled with phenomena and apparatuses. 
While this perspective offers a compelling critique of conventional research practices, its practical value for visualization research remains underexplored.
We apply the entanglement lens to examine and reframe the epistemic tensions that emerged in a longitudinal community-based visualization co-design project.
Our experience shows that the entanglement perspective not only provides a richer understanding of these tensions, but also helps transform them into generative opportunities for methodological and theoretical reflection.
Applying this lens enabled us to critically interrogate the language used in research, to develop a more nuanced understanding of visualization co-design, and to surface ``dark sides'' of conventional visualization design pipelines.
These contributions illustrate the practical value of embracing entanglement as an epistemological lens for visualization research. 
}


\graphicspath{{figs/}{figures/}{pictures/}{images/}{./}} 

\usepackage{tabu}                      
\usepackage{booktabs}                  
\usepackage{lipsum}                    
\usepackage{mwe}                       
\usepackage[dvipsnames]{xcolor}
\usepackage{xspace}
\usepackage{mathptmx}                  

\definecolor{artiCol}{HTML}{e492be}

\begin{document}



\maketitle

\section{Introduction}
\label{sec:intro}


The recent work of Akbaba et al.~\cite{Akbaba2025} introduces a thought-provoking epistemological framework for the visualization community: \emph{entanglements for visualization}. Grounded in feminist epistemology, the framework reconceptualizes how visualization knowledge is produced by arguing that \emph{knowledge artifacts} (e.g., data, visualizations, and insights) are not isolated entities, but are inherently entangled with the \emph{phenomena} they seek to understand and the \emph{apparatuses} through which they are produced. By foregrounding these relationships, the framework challenges conventional assumptions that treat visualization knowledge as objective, stable, and independent of the practices that generate it.

Although the framework is conceptually rich, to the best of our knowledge, Akbaba et al.'s case study remains the only published example that applies entanglement theory to examine a visualization project. More broadly, employing high-level epistemological theories to critically examine visualization practice and generate theoretical insights remains rare in the visualization community. In this paper, we contribute to this space by examining a longitudinal, community-based visualization co-design project through the lens of entanglement theory.

From 2023 to 2025, members of local creative communities on Vancouver Island, British Columbia, Canada, collaborated with a group of visualization researchers to co-create solutions with data science and visualization tools to explore data that is most relevant to the local arts sector. Conceived as a participatory initiative, the project sought to understand community data needs, collaboratively collect data, and design visualization tools that could support evidence-informed decision-making and policy advocacy.

As the collaboration unfolded, however, the project did not follow the envisioned trajectory. Instead, it exposed a series of epistemic tensions that challenged prevailing assumptions about visualization research and co-design practice. These tensions questioned how certain research terms were used, how co-design processes were conceptualized, and how conventional visualization pipelines framed the role of data. These tensions revealed mismatches between dominant epistemological assumptions in visualization research and the realities of community-based practice. 

Realizing that these tensions were rooted in the epistemological assumptions some of us were used to, we sought an alternative epistemological perspective. 
Akbaba et al.'s work on entanglements in visualization provided such a perspective: we came to understand these tensions as manifestations of entanglements in visualization. 
This shift in perspective enabled us to reframe the tensions as three productive opportunities for theoretical reflection. 
First, it prompts us to critically re-examine the language that visualization research uses. Second, it develops a more situated understanding of community-based visualization co-design by surfacing three fuzzy front ends. 
Third, it broadens critical visualization research by exposing limitations of conventional visualization pipeline models and highlighting how upstream data entanglements can shed light on the dark sides of visualization.

We also need to acknowledge several interrelated and entangled factors of this paper. First, this project was part of a bigger project with several interrelated research themes that while operating largely independently did have cross-project discussions and influences. Second, while members of this project are now more or less epistemologically in agreement, they started with very different stances; for example, some were starting as interpretivist and/or critical thinkers while some were consciously starting as positivists. 
To be able to most clearly tell the story and the impact of this project in this regard, this paper follows the changing viewpoints and awareness of the lead author, who started this journey from a positivist perspective. Some of the reflections outlined in this paper are informed directly by the first author's experiences and theoretical evolution, and should not be taken to represent the prior perspectives or experiences of all co-authors. 

By sharing our experiences and reflections, we demonstrate the practical importance of acknowledging the omnipresence of entanglements in applied visualization work. We hope this work motivates the community to move beyond conceptual engagement with entanglement theory and toward its active use for critically and constructively examining practice, rethinking established norms, and opening new possibilities for visualization design and research.

\section{Related Work}
\label{sec:rw}

In this section we briefly present the ongoing discussion on epistemology in visualization while highlighting Akbaba et al.'s work on entanglements for visualization.

\subsection{Visualization epistemology}
The predominant epistemology in visualization research is positivism~\cite{Akbaba2025}. According to \emph{The Oxford Dictionary of Philosophy}~\cite{blackburn2005oxford}, positivism is the view that believes:
\begin{quote}
    \textbf{the highest or only form of knowledge is the description of sensory phenomena.}
\end{quote}
As a form of empiricism, positivism holds that knowledge is derived from observable experience and that objective understanding can be achieved through empirical evidence. This epistemological stance has profoundly shaped visualization research, where data are often treated as objective representations of reality and where visualization is understood as a neutral medium for faithfully communicating or extracting knowledge from data.

However, this view is getting challenged. Researchers have started examining the emotional~\cite{Lan2024Affective}, rhetorical~\cite{Hullman2011}, and socio-cultural~\cite{Drk2013} dimensions of visualization, questioning established guidelines and assumptions~\cite{Akbaba2021Manifesto,kennedyWorkThatVisualisation2016}. 
Critical studies have interrogated dominant conventions, that include minimalist aesthetics, the maximization of the data-ink ratio~\cite{tufte1983visual}, and long-standing assumptions about perceptual efficiency and effectiveness~\cite{Mackinlay1986,Cleveland1984}. 
These critical works collectively argue that visualizations are not objective representations, but are shaped by rhetorical strategies and embedded within social and cultural contexts.

Kennedy et al.~\cite{kennedyWorkThatVisualisation2016} challenge this view by advocating for a social semiotic lens on visualization, emphasizing how conventional design choices can produce an ``aura of objectivity and neutrality.'' 
Hill et al.~\cite{hillVisualizingJunkBig2016} further demonstrate that judgments of what counts as a ``good'' or ``bad'' visualization are themselves socio-culturally situated. 
Peck et al.~\cite{peckDataPersonalAttitudes2019} show that interpretations of visualizations are shaped by viewers’ personal experiences, contextual knowledge, and relationships to the topic and data source. 
In politically charged contexts, Dörk et al.~\cite{Drk2013} propose a critical framework for information visualization that foregrounds power relations and introduces design principles emphasizing context and situatedness. 
Correll~\cite{correllEthicalDimensionsVisualization2019} emphasizes the ethical and political dimensions of visualization, arguing that design decisions are inseparable from questions of responsibility in knowledge production. 
Lee et al.~\cite{leeViralVisualizationsHow2021} illustrate how COVID-19 pandemic visualizations were mobilized by opposing groups to support conflicting narratives. 
Other visualization research also recognizes that visualizations can function as persuasive artifacts~\cite{Pandey2014Persuasive}, evoke and manipulate emotions~\cite{Lan2024Affective,blair2025emotions}, shape political attitudes~\cite{Holder2023Polarizing,Yang2023Swaying}, and reinforce existing biases~\cite{Xiong2022}.

Beyond empirical case studies, researchers have moved further to interrogate the theoretical foundations of visualization research directly, drawing inspiration from debates in the late twentieth-century ``science wars.'' 
Alternative philosophies of science---such as \textit{relativism} (the view that knowledge and its validity depend on perspective and context), \textit{constructivism} (the view that knowledge is produced by human choices and social negotiation), and \textit{feminism} (which, in this context, examines how power and privilege shape knowledge production, also known as feminism theory)---offer new lenses through which to reconsider visualization research and its epistemological commitments.
Within this broader turn, D’Ignazio and Klein’s \emph{Data Feminism}~\cite{DIgnazio2020} advances a feminist approach to data and visualization practice, highlighting how intersecting structures of race, gender, class, sexuality, ability, and age shape data production and interpretation. 
They also propose actionable principles for challenging and transforming inequitable power structures embedded in data practices. 
Akbaba et al.~\cite{Akbaba2025} further extended this line of inquiry by introducing entanglement theory to the visualization research community as an alternative epistemology for visualization, explicitly rethinking knowledge production as fundamentally entangled with the phenomena and apparatus. As we will describe in Section~\ref{sect:method}, our work is deeply motivated by Akbaba et al.'s conceptualization of entanglements for visualization, thus we briefly review their framework below together with its underlying theoretical foundations.

\subsection{Entanglements for Visualization}

The notion of \emph{entanglement} originates in quantum physics, where it describes non-classical correlations between quantum systems that cannot be explained independently of one another~\cite{Schrdinger1935}. In such systems, the state of each component cannot be fully specified without reference to the others, challenging classical assumptions of separability and independent existence.

As a theoretical physicist and a feminist scholar, Karen Barad extends this view to a philosophical framework in which object and observation, humans and non-humans, matter and meaning, are all mutually entangled and constituted. Barad calls this framework \emph{agential realism}~\cite{Barad1996}.
Within this framework, the notion of \emph{agential cut} is central. 
An agential cut is the situated and temporary boundary that delineates an object of inquiry from the broader field of phenomena in which it is entangled. 
In Barad's terms, making an agential cut refers to the use of a particular material apparatus to produce knowledge about a specific phenomenon~\cite{Barad1996}. 
These cuts are necessary for producing determinate knowledge, yet they are neither natural nor neutral. 
Each cut enacts a specific configuration of inclusion and exclusion, foregrounding certain relations while hiding others. 
In doing so, a cut establishes the boundary that determines what can be known and delimits what remains unknowable. 
Barad writes, \textit{``boundaries are not our enemies; they are necessary for making meanings, but this does not make them innocent''}~\cite[p.~345]{Barad2006}. Rather than accepting these boundaries as fixed or objective, Barad encourages us to interrogate and reconfigure them, arguing that shifting agential cuts can reveal previously obscured questions, particularly those concerning power~\cite{Barad1996}.

This epistemological shift resonates with Donna Haraway’s concept of \emph{situated knowledges}~\cite{2013womenScience}, which argues that all knowledge is produced from partial, embodied, and located perspectives. Rather than aiming for a ``view from nowhere,'' Haraway emphasizes that objectivity is always partial and constructed through specific positionalities, which shape what can be seen, known, and articulated.

Akbaba et al.~\cite{Akbaba2025} synthesize and extend these ideas by proposing entanglement as an alternative epistemology for visualization research. 
Drawing on Barad's agential realism, they revisit three foundational concepts in visualization research: data, visualizations, and insights. 
Instead of viewing them as discrete, stable entities, they argue that each should be understood as a situated, momentary, and entangled knowledge artifact, produced through specific relations among phenomena and apparatuses. 
They further demonstrate how this perspective reframes these foundational concepts, then present a series of provocations that illustrate the generative possibilities of adopting an entanglement perspective.
Akbaba et al. further illustrate this reframing through examples such as implicit error, showing how what is treated as ``error'' is itself entangled with contextual assumptions and methodological choices rather than being an objective deviation from truth.



\section{Clarification of Terms and Pronouns}
We follow the use of two key concepts in Akbaba et al.' work~\cite{Akbaba2025}:
\textit{knowledge artifact} and \textit{epistemic tension}.

According to their definition, knowledge artifact refers to \textit{``the static, momentary, bounded representation''} that is produced by \textit{``the interaction between an apparatus and a phenomenon.''} 
To this extent, nearly all knowledge that humans possess are knowledge artifacts that are produced under a specific and situated choice (namely, an agential cut); a choice that determines what the knowledge is about (phenomenon) and how the knowledge is gained (apparatus). 
Consequently, this choice also determines what is not included in the knowledge and excludes other approaches that could be used to produce the knowledge. 
For the visualization community, concepts like data, visualization, and insight are all knowledge artifacts.
Akbaba et al. used the term epistemic tension without explicitly defining it. 
Here we adopt the term to describe the intellectual tensions that arise when the dominant epistemology underlying a knowledge-production process is insufficient to guide or explain what happens in practice.


This paper is based on a three-year, community-involved, co-design project, which included a changing number of researchers and arts community members. 
To acknowledge this, the pronouns \emph{we}, \emph{us}, and \emph{our} refer to the authors of this paper. 
\emph{The team} refers to all researchers and community members who participated in this project. 
\emph{Visualization researchers} and \emph{community members} are used to highlight the particular agency, perspective, and contribution from visualization researchers or from arts community members. 
We also acknowledge that neither the visualization group nor the arts group are unified in the way they think. They both are composed of individuals who have differing epistemological views--- see the positionality statements in Section ~\ref{sec:postionality}.

Lastly, as the project progressed, language itself became a key perspective for reflection. 
The language we had adopted during the early stages of the project (e.g., grant proposal and project plan) was a compromise between differing epistemological views and the accepted vocabulary norms for describing visualization research. 
As we discuss in Section~\ref{sec:results}, we came to a more unified recognition that several terms like \emph{user}, \emph{user evaluation}, \emph{data empowerment}, and \emph{data clinics} carried false assumptions and biases. 
Nevertheless, rather than retrospectively rewriting the project through our current understanding, we preserve these terms as historical artifacts of the collaboration and critically examine their implications later through the lens of entanglement theory.

\section{Context for this work}
\label{sec:background}

This discussion presents the results of an interdisciplinary team's reflection and introspection over the course of three years of a visualization co-design project.
From 2023 to 2025, a team of artists, cultural leaders, and researchers aimed to uncover narratives embedded in data that are most relevant to the local creative community, and to co-design potential solutions for improved data accessibility and community empowerment. 
This project was initiated by the local arts communities on Vancouver Island, British Columbia, Canada. Vancouver Island is home to a vibrant and well-established arts and culture sector. The region has a diverse network of artists, cultural workers, and grassroots organizations, with arts and cultural activities contributing significantly to both local identity and the regional economy, with at least one in twenty people involved in the arts and culture sector. According to the 2021 \textit{Arts Impact Study}, the arts and culture sector on Vancouver Island generated an estimated economic impact of over \$900 million~\cite{Nordicity2021}.

However, the economic support received by the creative community is disproportionate to their contributions. This situation worsened when the COVID-19 pandemic disrupted the sector, intensifying existing precarity within many arts organizations and communities. Emerging from the social and economic disruptions of the pandemic, members of the arts sector began mobilizing to re-imagine how they could sustain, understand, and advocate for their future. Motivated by the findings of the 2021 \textit{Arts Impact Study}~\cite{Nordicity2021}, community members recognized that collecting, analyzing, and communicating data that is relevant to them can help better understand and communicate the lived realities of their own community, facilitate more effective lobbying, and ideally promote policy changes and support long-term planning.

Unfortunately, existing data infrastructures were often externally controlled and lacked accessibility, which limited the ability of communities to meaningfully engage with (not even mention to benefit from) their own data. This motivated the creative community to seek more active participation in the collection, interpretation, and communication of data relevant to their own needs and realities.
It is in that context that the creative community reached out to data and visualization researchers to collaborate on exploring how data practices and technology tools could support their needs.
Central to this collaboration was a shared recognition of the needs to 
1) understand and collect relevant data that ensures inclusive representation of diverse groups, 
2) analyze these data to generate meaningful insights, 
and 3) communicate findings in ways that effectively support public dialogue and policy advocacy.
Together, these motivations positioned the project as an effort to transform community data practices from passive reliance on external systems toward active, community-led data empowerment.

\subsection{Methodology}
\label{sect:method}

This paper does not report the findings of a conventional empirical research study. Although the project involved sustained collaboration between researchers and community members over three years, it was never conceived primarily as a research study. In fact, as shown in Section~\ref{sec:codesign-lang}, community partners explicitly emphasized that \emph{they did not wish to be positioned as research participants}. Instead, they sought a collaboration that was community-driven, community-led, and community-benefitted, with the primary goal of supporting the needs and aspirations of the Vancouver Island Arts sector.

As a result, the project was not organized around the systematic collection of data for later analysis. Many activities that might ordinarily be treated as research materials---such as workshop discussions, meetings notes, design sessions, and community events---were conducted to support the project's practical objectives rather than to generate data for research purposes. Consequently, this work should not be understood as an empirical or an autoethnographic study~\cite{Kaltenhauser2024}, despite sharing certain characteristics with practice-based and reflexive research.

Rather, this paper presents retrospective and theory-informed reflections on three years of collective engagement, including workshops, interviews, public demonstrations, and more than 100 team meetings involving researchers and community partners. Throughout the project, team members continually reflected on both successes and challenges as they sought to advance the project's goals. These ongoing discussions generated a growing awareness that many of the difficulties encountered were not simply technical, methodological, or organizational in nature. Instead, they reflected deeper underlying tensions that stemmed from the assumptions embedded in the epistemology that many of the researchers held, i.e., they are epistemic tensions.    

This realization led us to use entanglement theory as a conceptual lens for examining and reframing the collaboration. 
Through this lens, we came to view the project as a rich example of how visualization design unfolds within complex sociotechnical assemblages, where data, visualizations, insights, people, power, and understandings of the project itself are dynamically entangled and continually co-produced.

Guided by this perspective, we engaged in a retrospective process of collective reflection. Through a series of discussions and iterative sense-making sessions steered by the first author, we revisited the whole journey of the project, identified three main epistemic tensions, and reframed the project through interrogating knowledge artifacts behind each tension. These reflections were refined through constant discussions among the authors. 
We further considered the implications of these themes for both visualization research and visualization practice.
\section{Epistemic Tensions that Unfolded in The Project}
\label{sec:codesign}

To pursue the project goals we mentioned in Section~\ref{sec:background}, 
the team adopted co-design~\cite{Sanders2008} as the main methodology for this project. Co-design emphasizes the integration of domain knowledge and iterative prototyping through collaboration between diverse stakeholders. Given the community-based nature of the project, the team regarded co-design as an appropriate methodological foundation. 
The team envisioned the project as a relatively structured process with three stages: 1) collaboratively identifying community needs and collecting relevant data, 2) developing visualization tools to support exploration and sensemaking, and 3) deploying and evaluating these tools with community members.

However, as the collaboration progressed, the team encountered a series of tensions that emerged from the interaction between the assumptions embedded in epistemology of visualization researchers and the realities of visualization co-design and of collaborating with a community whose priorities, experiences, and ways of knowing did not always align with those theories. 



\subsection{Tension Regarding Research Language}
\label{sec:codesign-lang}

The first stage of the project focused on developing a foundational understanding of the data needs and realities of Vancouver Island's arts ecosystem. The initial proposal envisioned constructing a dataset that would capture economic and social dimensions of the region's creative sector. This dataset would provide the basis for subsequent data exploration, visualization design, and insight generation, ultimately supporting policy advocacy, funding allocation decisions, and broader goals of community data empowerment.

The  proposal recognized an asymmetry of expertise among community members and researchers. While some community members had started to actively engage with data, largely community members' expertise was in lived experience about how existing data was not benefiting them and how some of the data that might benefit them was now considered proprietorial.
Neither the community members nor the researchers knew what data would be most meaningful or actionable, nor if it could be collected. While the visualization researchers possessed expertise in data collection, analysis, and representation, they lacked the situated domain knowledge necessary to determine which data were most relevant to the lived realities of the creative sector.

To bridge this gap, the researchers initially proposed a series of data activities including various types of data workshops, 
aimed at familiarizing people with working with data 
and building mutual understanding between researchers and community partners. These activities were intended to help community members explore how they think about, use, interpret, and act upon data within their everyday practices. 
More specifically, the workshops sought to address questions such as: What data are most relevant to the community? What data are already available? What data exist but remain inaccessible? What additional data should be collected, and through what means? 
By holding workshops around these questions, the project aimed to surface community priorities, identify meaningful opportunities for data collection, and establish a shared foundation for subsequent co-design activities.

Surprisingly, the first tension emerged before even the project's planned activities had begun, when visualization researchers proposed data workshops as just described to identify data needs and develop data understanding in order to achieve the goal of data empowerment.
However, those initially suggested workshops made use of toys and craft objects such as wooden beads, string and tape, which the creative community said were demeaning for them as artists. 
They pointed out that they might be successful with WEIRD people (White, Educated, Industrialized, Rich and Democratic), but that they were not acceptable for members of the creative community.
In addition, from the perspective of the visualization researchers, terms like data empowerment and data workshops seemed to be straightforward enactions of the descriptions of the project's objectives. However, community members interpreted them differently and
several collaborators expressed discomfort with the language used in the proposal. 
Terms such as \emph{data empowerment} implied that community members were deficient in power as well as data literacy or agency and therefore required intervention from external experts. 
Likewise, we used the phrase \emph{data clinics} in our grant application, which we now realize suggested a relationship in which researchers would diagnose and address problems on behalf of the community. 
Our proposal did include terms such as \emph{users} and \emph{evaluation}, which are still commonly employed in visualization research, even though some of our research members have long questioned the use of words like users\cite{bradley2015gendered},
This language use was increasingly questioned and interrogated during the project. 
The implied passivity of community members embedded in these terms implicitly conveys structures of power and agency, regardless of the research team's intentions. 
They shaped how community members understood the project and their place within it, leading community members to state: \textit{``We do not want to be your study participants.''}
This forms the first epistemic tension:
\begin{quote}
\textbf{First epistemic tension: \textit{the tension between the assumed neutrality of research language, and the implicit structures of power and agency it encodes.}}    
\end{quote}


\begin{figure}[t]
    \centering
    \includegraphics[width=1.0\linewidth]{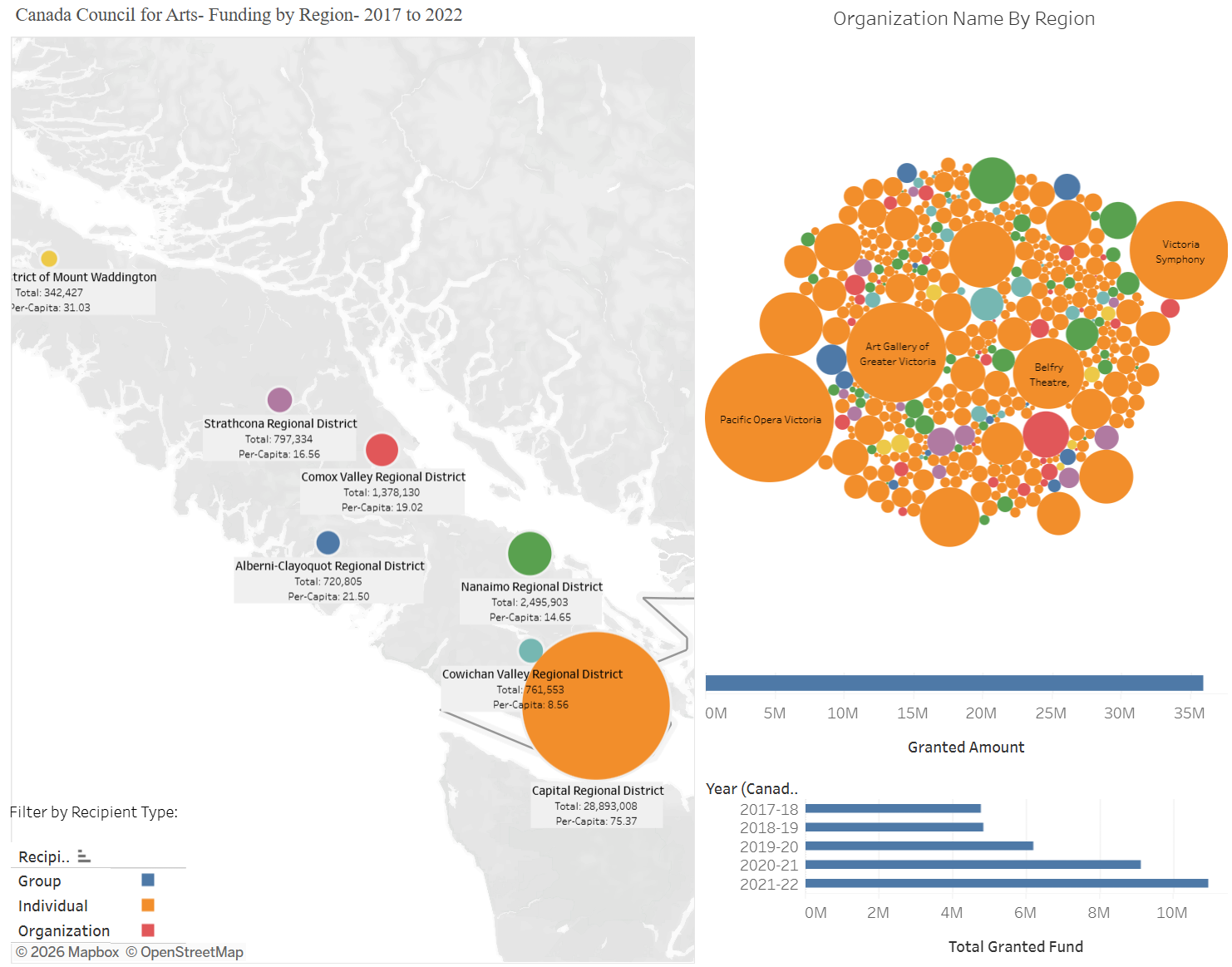}
    \caption{An example of the preliminary dashboard prototype developed in Tableau during the early stages of our co-design process.}
    \label{fig:tableau}
\end{figure}

\subsection{Tension Regarding Co-design Methodology}

Several of the initial tensions around language and activities were resolved by moving away from data workshops and developing Data Makers Residencies, where artists used their own materials in their own studios.
This resulted in numerous explorations of data making, data art, and data in the community as part of the crossovers between art and technology. 
However, we encountered a second tension when we moved to visualization co-design -- the co-design of visualizations for existing data that has been gathered about the arts.

In this part of the project, the visualization researchers had originally planned to collect a new dataset. However, they soon realized that the arts sector could potentially benefit from many different types of data and that collecting data would be another multi-year project. 
In order to begin the project from a concrete starting point, we decided to work with existing datasets, hoping that engaging with available datasets would allow us to explore what kinds of insights visualization could support, while also identifying what data might be missing. 
We started with the dataset published by the \emph{Canada Council for the Arts}, Canada's national public arts funder and the primary source of federal funding for artists and arts organizations. As part of its public accountability efforts, the Council proactively publishes records of all grants and prizes awarded on its website~\cite{CanadaCouncilArt}.
The dataset includes basic information about each funding recipient, including the recipient's name, organization type, location, and awarded amount.

Based on this dataset, one visualization researcher began constructing visualizations in Tableau through regular co-design sessions with community members. 
Early meetings focused primarily on representation: how should funding be encoded, what metaphors would resonate with the community, and how could the visualizations reflect the geographical character of Vancouver Island's arts ecosystem? 
The team experimented with various designs, including representing funding through bubbles of different sizes and superimposing visualizations onto a map of Vancouver Island. 
After several weeks, a collection of Tableau dashboards had been developed (Fig.~\ref{fig:tableau}). 
Although many community members had little previous experience with interactive visualizations, the prototypes generated considerable enthusiasm and sparked lively discussions about both the data and the designs.

For a brief period, it seemed that the project had found its direction. Yet, each design decision unexpectedly opened another set of unresolved questions. Community members envisioned advanced interactions that Tableau could not readily support, prompting the team to reconsider the technological platform. 
As community members explored the visualizations, they also began asking questions that could not be answered with this dataset. 
For example, the dataset only includes funding recipients from the Canada Council for the Arts, capturing awards at the federal level, but not awards at the provincial or municipal levels, resulting in a partial view of the broader arts funding landscape. 
The visualization, rather than simply communicating the data, exposed the limitations of the data. This realization led the team to explore and identify a richer dataset from the Canada Revenue Agency, Canada's federal tax administration department. 
Meanwhile, another visualization researcher joined the project with the technical mission to develop a custom interactive visualization capable of supporting both the larger dataset and the increasingly sophisticated interaction requirements.

Rather than simply replacing one dataset with another, however, this decision returned the project to questions that the team had assumed were already resolved. The team revisited the structure, meaning, and potential uses of the new dataset. Community members drew upon their domain expertise to identify meaningful aspects and relationships, while visualization researchers simultaneously reconsidered visual encodings and interaction techniques. New interaction ideas reshaped discussions about which aspects of the data were most meaningful; new understandings of the data, in turn, inspired new visualization designs. Progress in one aspect of the project repeatedly unsettled decisions that had already been made in another. Instead of advancing with a sequence of stages, the project cycled continuously between understanding the data, designing interactions, and refining visual representations.

\begin{figure}[t]
    \centering
    \includegraphics[width=1.0\linewidth]{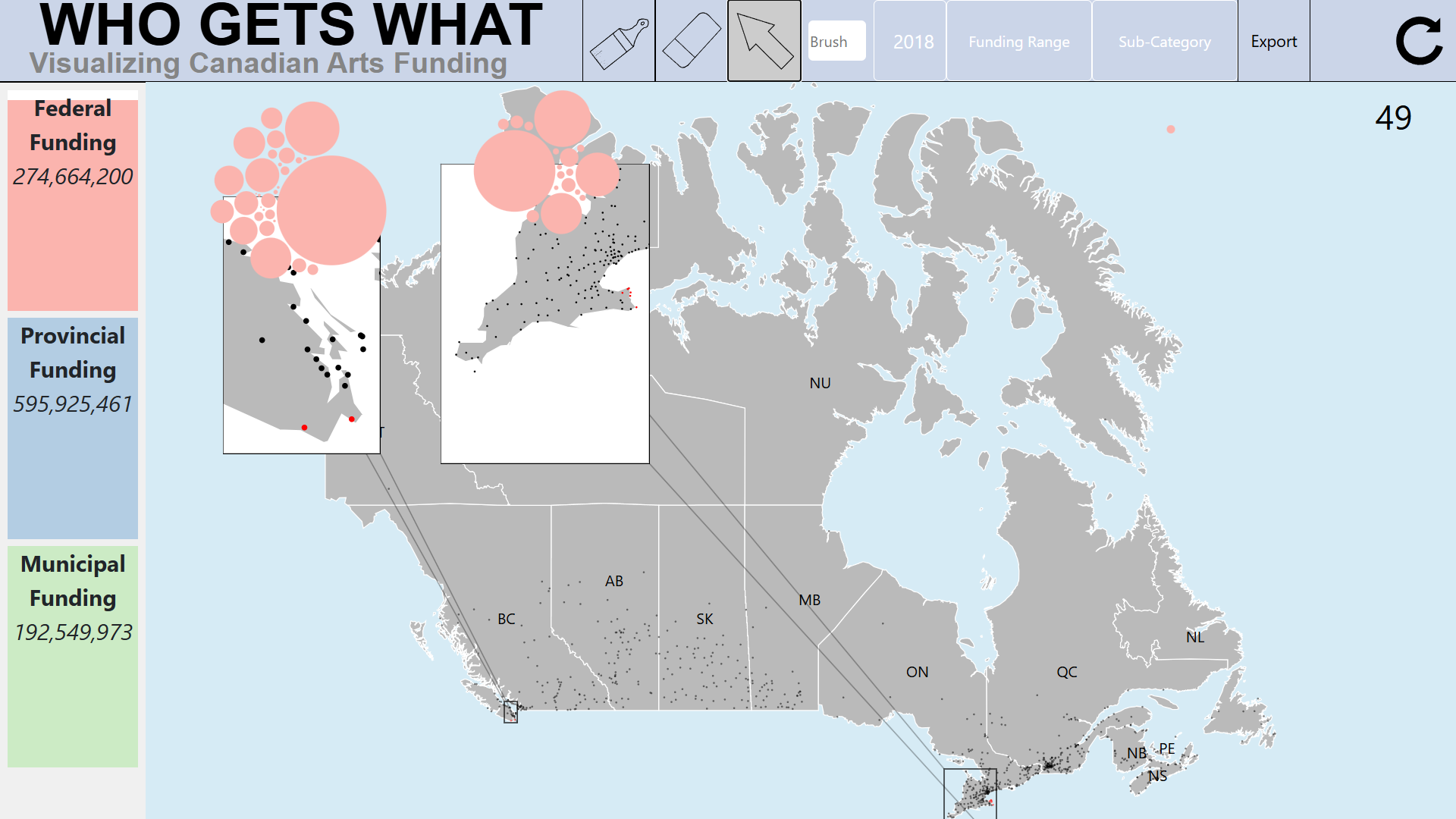}
    \caption{The ``data painter'' developed to support larger dataset and better interaction.}
    \label{fig:datapainter}
\end{figure}

\begin{figure}
    \centering
    \includegraphics[width=\linewidth]{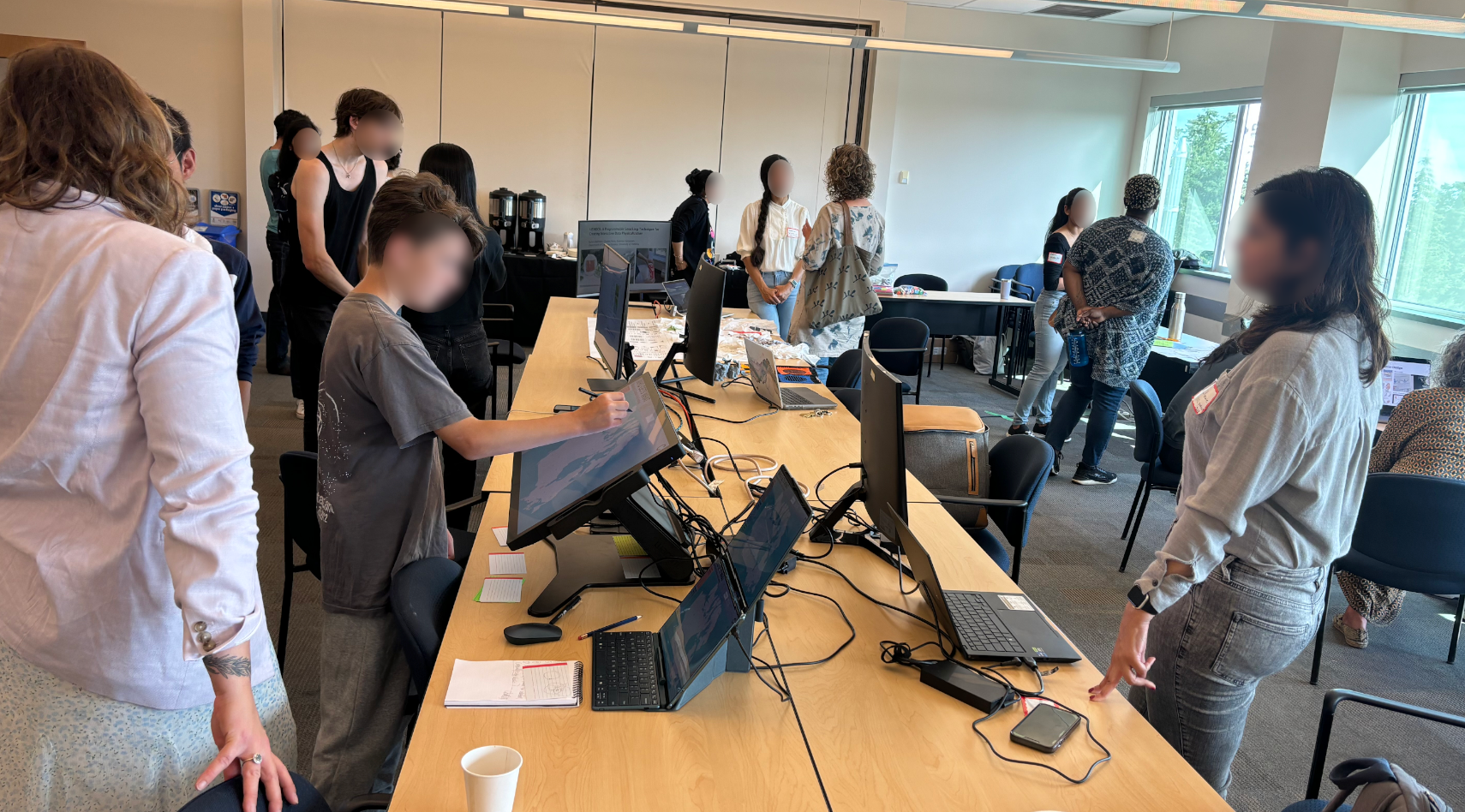}
    \caption{Photo taken at the technology festival where community members interacted with novel research projects, including the ``data painter''.}
    \label{fig:techfest}
\end{figure}

Months of these iterative and intertwined explorations eventually culminated in the development of \emph{Data Painter}, a web-based visualization that lets people ``paint'' the landscape of Canada's arts funding distribution using a virtual brush (Fig.~\ref{fig:datapainter}). Confident that the visualization was mature enough, the team organized a technology festival to introduce the tool~\footnote{\url{https://www.codesignexplore.com/}}, along with a dozen other novel research projects from the larger project, to a broader audience of artists, community organizations, and researchers (Fig.~\ref{fig:techfest}). The aim was to demonstrate the visualization and to gather feedback for the next round of refinements.

The  technology festival was well received, and participants provided numerous constructive suggestions for improving the tool. Once again, it appeared that the project had reached a new milestone and that the next step would simply be another iteration of the visualization design.

Instead, the discussions following the showcase reopened an even more fundamental question. In conversations between visualization researchers and arts community members, attention shifted away from the visualization itself and back to the dataset. 
Does this dataset actually represent the lived realities of arts community? 
What dimensions of artistic practice are absent? 
How might these absences shape or bias the insights generated through the visualization? 
It turns out that this dataset---although the most comprehensive one the team found---is still incomplete and biased. 
The nature of the dataset (i.e., collected by revenue agency) means it only includes the information of artists and arts organization who were funded (as only then they would appear in the tax report), while it excludes those who were not. 
In short, the dataset only represents the ``bright'' side of the living realities of arts communities. The team needed a more comprehensive dataset that could reveal both the ``bright'' and ``dark'' side.

After more than two years of collaboration, the team found itself returning to the very questions that had initiated the project: what data should be collected? What realities should the visualization represent? Whose understanding of the arts ecosystem was being visualized?

The team had anticipated that a community-based co-design process would not unfold in a strictly linear manner. The original project plan already incorporated opportunities for feedback and iteration, acknowledging that design decisions would evolve through ongoing collaboration with community partners. 
Nevertheless, some team members still implicitly understood visualization co-design as progressing through a sequence of distinguishable stages: developing an understanding of the data, designing visualizations, evaluating prototypes, and refining the resulting tools. Iteration was expected, but, by some, it was assumed to occur largely within each stage before the project moved on to the next.

\begin{figure}
    \centering
    \includegraphics[width=\linewidth]{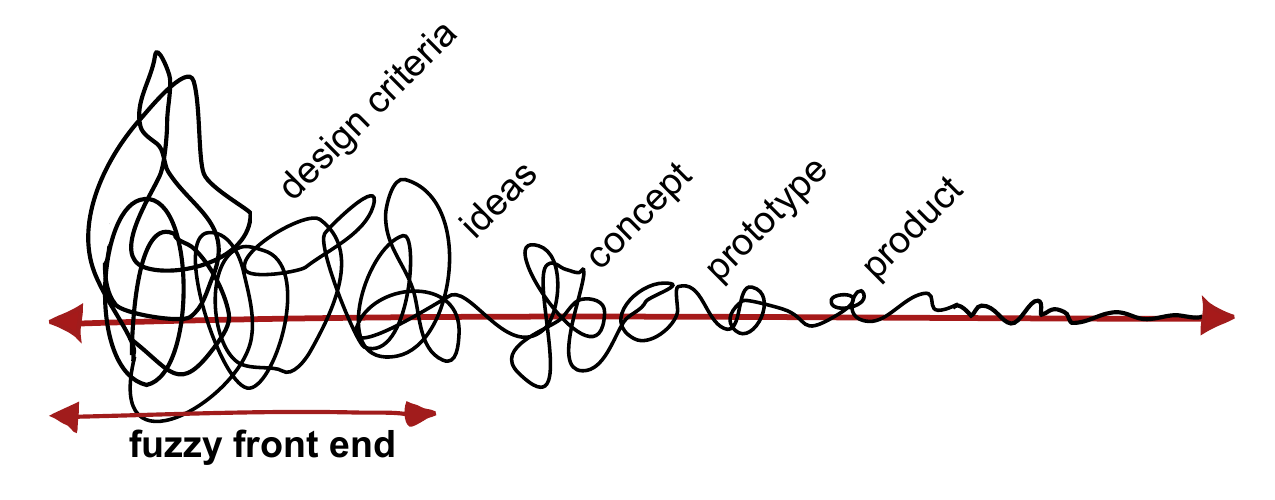}
    \caption{The co-design process illustration that highlighted the fuzzy front end. Redrawn from Sanders \& Stappers' paper~\cite{Sanders2008}.}
    \label{fig:fuzzyfront}
\end{figure}

Our experience challenged this understanding. Rather than progressing from one stage to another, the project repeatedly returned to questions that some members of the team believed had already been resolved. 
In retrospect, we realized that the team had never truly left what design research describes as the \emph{fuzzy front end} of co-design (Fig. \ref{fig:fuzzyfront}), the phase in which problem framing, stakeholder needs, design goals, and opportunities remain fluid and continuously negotiated and re-negotiated. Although this concept is well recognized in design literature, it is often treated as a preliminary stage. 
Our experience suggested something quite different. Throughout more than two years of collaboration, the fuzzy front end persisted. Data understanding, visualization design, interaction design remained deeply intertwined, continually reshaping one another rather than stabilizing into successive phases.
Here, we found ourselves confronting the second epistemic tension: 

\begin{quote}
\textbf{Second epistemic tension: \textit{the tension between the abstract conceptualization of the co-design process, and its real-world enactment in a community-based visualization project.}}    
\end{quote}

\subsection{Tension Regarding the Visualization Pipeline}

While we were grappling with the recurring return to the fuzzy front ends of co-design, we also started to pay attention to something we referred to as the ``dark side'' of data. As described earlier, recognizing the biases and incompleteness of the dataset prompted us to reflect on what such absence implies for visualization practice. This led to a growing awareness that the arts funding data we were working with, as well as the visualization that we had created, did not present the full picture of real arts ecosystem--it constructed a partial view of it.

For the visualization researchers, this led to the realization that any incompleteness or bias in the dataset propagated through the visualization pipeline, shaping the representations of the arts sector. 
The visualizations, while coherent and informative, reflected the visible domain of funding allocation. 
This perspective systematically excluded many forms of artistic labor and precarity: artists who never received funding, organizations whose applications were repeatedly rejected, and long-term cultural work sustained without formal financial recognition.

More critically, this absence was not neutral. 
Because funding systems already privilege certain institutions and practices, marginalized and underrepresented groups were disproportionately absent from the dataset. 
As a result, these groups were also at risk of being absent from the visualizations that claimed to represent the arts landscape. 
This raised an uncomfortable question for the team: \emph{do such visualization systems merely fail to represent underrepresented communities, or do they actively reproduce and reinforce their marginalization by rendering them invisible within data-driven representations?}

\begin{figure}[t]
    \centering
    \includegraphics[width=\linewidth]{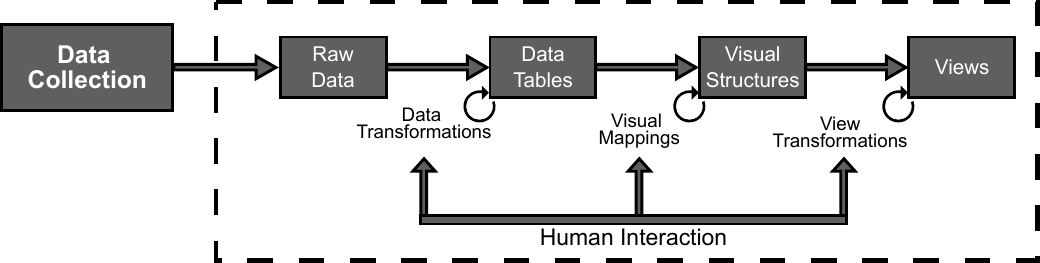}
    \caption{Our experience revealed that ``dark sides'' of data propagate throughout the visualization pipeline, rendering subsequent stages ``dark'' as well. Redrawn and extended from Card's reference model~\cite{card2009information}.}
    \label{fig:darksidesModel}
\end{figure}

These concerns prompted a broader reflection on the assumptions embedded in the visualization pipeline (e.g., the model within the dashed rectangle in Fig.~\ref{fig:darksidesModel}). 
Visualization research often implicitly treats data as a given input--often assumed to be objective, neutral, and reliable--while positioning the task of visualization as transforming data into insights through effective encoding. 
In this framing, responsibility for data quality is largely external to visualization design. 
However, our experience suggests that this separation is untenable in practice. 
When data is incomplete, biased, or structurally uneven from the outset, visualization does not merely miss some information--it also risks amplifying the very absences and distortions embedded in the dataset. 
If the data encodes structural inequalities, then the visualization becomes a mechanism that can legitimize and reinforce those inequalities. 
This led us to the third epistemic tension: 
\begin{quote}
\textbf{Third epistemic tension: \textit{the tension between visualization pipeline models that assume data to be objective pre-existing inputs, and community contexts where data is inherently situated and partial.}}    
\end{quote}

\section{Reframing Epistemic Tensions}
\label{sec:results}

The epistemic tensions described in the previous section persistently challenged the commonly held stance of positivism of some of the team's visualization researchers and of the larger international visualization research community~\cite{Akbaba2025}. 
This perspective (initially shared by the first author) thus was the underlying stance adopted at the outset of this visualization project. 
These tensions revealed difficulties that could not be adequately explained by assumptions of objectivity or empirical observation alone. 
Entanglement theory, however, provided a different lens through which to reflect on and reframe these tensions.

\subsection{Word and World}
The first tension that emerged in the project concerned the relationship between language and knowledge. 
At the beginning of the collaboration, some members of the visualization group regarded terms such as \emph{users},  \emph{data empowerment}, and \emph{data clinics} as neutral descriptions of the project's objectives. 
However, community members stressed that these terms conveyed assumptions about expertise, authority, and the respective roles of researchers and community members. 
Before any data had been collected or any visualization designed, language had already shaped how the collaboration was understood and enacted.

Entanglement epistemology offers a constructive way of understanding this tension. Rather than treating language as a neutral medium for describing reality, it allows us to understand language as a knowledge artifact bounded by the historical, institutional, and cultural contexts in which it is produced, and entangled with broader structures such as disciplinary conventions, politics,  and capitalism. 
From this perspective, words are not merely tools for representing the world; they participate in producing (and twisting) it~\cite{norman2006words}. 
Terms such as \emph{data empowerment} and \emph{data clinics}, as well as commonly used vocabulary in HCI and visualization research such as \emph{users}, \emph{subjects}, and \emph{user evaluation}, are not neutral technical descriptors. 
Rather, they are historical knowledge artifacts that embody particular (most of the time inaccurate and outdated) assumptions about expertise, agency, and authority.

Some of these assumptions become especially visible in our project in retrospect. 
For example, terms like \emph{users} or \emph{subjects} may implicitly position the people that researchers work with as passive recipients of systems or knowledge, rather than active contributors to its production. 
By inheriting this vocabulary without careful scrutinization, researchers may also inherit the assumptions embedded within it, which in turn shape how relationships with communities are framed and enacted. 
In fact, these terms become so normalized within research practice that the assumptions they carry often become invisible to those who use them, a realization that, in our case, only became apparent through sustained engagement with community partners.

This has important implications for visualization research. If language constitutes (rather than merely describes) realities, then results and reflections from research must extend beyond methods to include the conceptual and linguistic categories through which research is framed. 
Choosing words is therefore not only a matter of communication or inclusivity; it is also an epistemic and ethical practice that shapes relations of power, forms of participation, and what kinds of knowledge can emerge.
A straightforward response might be to replace problematic terminology, for example, using ``people'' or ``collaborators'' instead of ``users,'' using ``participants'' instead of ``subjects.'' 
Such recommendations have been discussed in the research community~\cite{vandenHoonaard2008, norman2006words, Bannon2011}. However, our experience suggests that changing terminology alone is insufficient (using alternative words for ``users'' might not be sufficient to shift the community's impression that they are passive entities of an empirical investigation).  
What is also required is to recognize and shift epistemological assumptions that cause these problematic terms. Mitigating this tension therefore involves not only revising vocabulary, but also critically examining these implicit assumptions about agency, value, and power that are embedded in everyday research language.

\subsection{Three Fuzzy Front Ends of Visualization Co-design}
The second tension concerning the co-design process prompted us to reflect on what it means to conduct co-design in practice. 
While the literature recognizes co-design as iterative, iteration is often still framed within a broadly staged progression from problem definition to design and evaluation. 
Our experience suggests that in community-based visualization projects, these activities do not unfold as separable phases but remain fundamentally intertwined throughout the collaboration.

Viewed through the lens of entanglement, this tension reflects the limitations of treating co-design as a universal and stable methodology. 
Instead, co-design itself can be understood as a knowledge artifact, shaped by the historical and material conditions from which it emerged. 
Originating in participatory design practices in Northern Europe~\cite{Sanders2008}, co-design was initially developed to improve industrial systems by involving workers in design processes. 
As it has been adopted by service design~\cite{steen2011benefits}, HCI~\cite{muller2002participatory}, and visualization research~\cite{Drk2020}, its theories and practices have inevitably become entangled with the distinct goals, paradigms, and contexts of each domain. 

This perspective shifts the second tension into a more situated understanding of co-design in visualization research. Rather than following a staged process with a single ``fuzzy front end,'' our project revealed multiple, recurring fuzzy front ends that emerged throughout the collaboration. Specifically, we identified at least three interrelated fuzzy front ends: those concerning \emph{data} (i.e., activities related to data collection, interpretation, and wrangling), \emph{representation} (i.e., activities related to design the visual structures of visualizations), and \emph{interaction}  (i.e., activities related to design interaction techniques of visualizations). As narrated in Section~\ref{sec:codesign}, these fuzzy front ends continually shaped one another as the project unfolded rather than being resolved sequentially.

For example, in the early stages of the project, we experimented with bubble-plot representations in Tableau using a readily available dataset to rapidly prototype ideas. However, the limitations of both the dataset and the off-the-shelf visualization tool quickly became apparent. Decisions made to facilitate early prototyping constrained subsequent discussions about data collection and interaction, while emerging requirements from later stages repeatedly prompted us to revisit these earlier choices. Instead of progressing linearly, questions about data, representation, and interaction remained interdependent throughout the collaboration (a more detailed and visual account of these three fuzzy front ends can be found in our previous work~\cite{wei2026fuzzy}).

In this case, shifting epistemological perspective does not solely help explain the tension the team encountered; it also actively transforms it into a generative site of knowledge production about co-design in visualization research.

\subsection{Dark Sides of Visualization}

The third epistemic tension made us reconsider one of the most fundamental assumptions in visualization research: the visualization design pipeline(s). 
From an entanglement perspective, the well-established visualization pipeline can be understood as a knowledge artifact. 
It is bounded by the priorities and assumptions of the visualization community, placing primary emphasis on visual representation and interaction design while largely treating data as a given input. 
However, in reality, a visualization is inevitably entangled with the input data. 
Our experience revealed that visualization design is inseparable from the data on which it is built. 
Also,when data is incomplete, biased, or systematically excluding certain communities, these problems propagate throughout the pipeline and may even be amplified by visual representation and interaction, as shown in Fig.~\ref{fig:darksidesModel}.

The third epistemic tension extends our understanding of when and how visualization design can go wrong~\cite{correll2017black}. 
Visualization researchers have studied and documented ways in which visualization can mislead, misinform, or undermine understanding (e.g., ``chartjunk,''~\cite{tufte1983visual,hillVisualizingJunkBig2016} ``misleading visualization,''~\cite{Szafir2018,huff2023lie,Cairo2019,Lisnic2023} ``design flaw,''~\cite{Lan2025} ``visualization pitfalls,''~\cite{Bresciani2015} ``misinformative visualization,''~\cite{Lo2022} and ``deceptive design''~\cite{Lauer2020}). 
While all these concepts focus on how design choices can distort interpretation and impair understanding, our experience suggests that visualization failures may also arise earlier in the entangled processes through which data is collected, interpreted, and filtered. 
The entanglement perspective shifts the focus of critique from visualization artifacts alone to a broader network of data practices and infrastructures that make visualization possible.

Inspired by this perspective, we propose a broader conceptualization: the \emph{dark sides of visualization}, referring to the misleading, exclusionary, inequitable, or otherwise harmful consequences that visualization systems may produce. 
Existing concerns such as misleading visual encodings and deceptive representations constitute one category of the dark sides of visualization, which we term \emph{design-related dark sides}. 
However, we argue that an equally critical yet underexplored category lies upstream: \emph{data-related dark sides}, where harms arises from missing data, biased data, structural invisibility in collection, and other upstream conditions that precede visualization design. 
Because these conditions remain entangled with every subsequent stage of visualization design, they ultimately shape not only what becomes visible, but also what remains invisible, unquestioned, and beyond action.

Similar to the second epistemic tension, here shifting epistemological perspective enables us to transform the tension around visualization pipeline to a generative site for expanding the understanding of how visualization can result in harms.

\section{Call for Awareness of Alternative Epistemologies}

This work delineates our experience of employing an alternative epistemological lens to examine, reflect on, and reframe a community-based visualization co-design project. We share this experience because the three-year collaboration was intellectually challenging, 
and we think our experience is not a unique case. 
As visualization research increasingly engages with community partners, interdisciplinary collaborators, and socially situated problems~\cite{Kerzner2019,Thompson2023, Khowaja2022, Snyder2020}, similar epistemic tensions are likely to become more common.

Importantly, most of the challenges we encountered did not stem from technical limitations, such as algorithms, hardware, design or programming capabilities, but from the implicit assumptions held about what constitutes rigorous scientific research and how knowledge should be produced. 
For example, the tension around language could not have been resolved simply by using some synonyms or employing a different communication tool. Instead, it reflected deeper assumptions about objectivity and agency in knowledge production. These assumptions, largely inherited from the positivist traditions that have long shaped visualization research, are so deeply embedded in many of our members and our practices that we rarely recognize them explicitly. This project forced us to notice these hidden assumptions, and the lens of entanglement theory enabled us to interrogate these assumptions.

Our purpose, however, is not to argue that positivism should be abandoned, nor that entanglement theory provides the definitive solution to epistemic tensions in visualization research. 
Instead, the message we aim to convey is that researchers should be actively aware of the epistemological commitments underlying their work and the assumptions those commitments inevitably carry. 
As Thomas S. Kuhn argues in \textit{The Structure of Scientific Revolutions}, scientific inquiry is always guided by ``\emph{some implicit body of intertwined theoretical and methodological belief that permits selection, evaluation, and criticism.}''\cite[p. 28]{kuhn1970structure}. 
Making these implicit assumptions and beliefs explicit enables researchers to better interpret and situate research results, and, when it occurs, to engage epistemic tensions as opportunities for reflection rather than obstacles to overcome.

We also encourage visualization researchers to become more aware of alternative epistemological perspectives and to remain open to engaging with them when appropriate. 
Such openness does not mean abandoning established research traditions. Rather, it expands the repertoire of conceptual lenses available, enabling researchers to interpret epistemic tensions differently and, in some cases, transform those tensions into opportunities for new forms of understanding, as our experience illustrates.

Beyond individual research projects, we believe this awareness also has implications for the visualization research community more broadly. As the field increasingly embraces qualitative, participatory, and critically oriented research, it will inevitably encounter a wider diversity of epistemological perspectives. Similar shifts have already emerged within the HCI community, where non-traditional methodologies or those adapted from other domains have gained recognition as legitimate forms of inquiry, such as autoethnography~\cite{Kaltenhauser2024}, autobiographical design~\cite{Neustaedter2012}, and design fiction~\cite{Bleecker2022}. Supporting epistemological plurality in visualization research may therefore require not only new research practices but also more reflective reviewing criteria that recognize and welcome contributions grounded in different ways of knowing.











\section{Positionality Statements}
\label{sec:postionality}
In many ways, this entire paper acts as a positionality statement, documenting our collective examination and reflection on the epistemological foundations of our work. However, a singular voice cannot capture the complex makeup of our team. Our co-design team includes two senior visualization professors, two postdoctoral researchers, one PhD candidate, two creative community artists from Vancouver Island, and one trans-disciplinary creative artist/technologist/storyteller. We held different, and evolving, views about knowledge production over the course of the project. We therefore share individual statements below to honor and disclose this diversity in background and perspective.

\textbf{Wei Wei}: I am an international PhD candidate at the University of Victoria, BC, Canada. Having received 10 years of scientific and academic training in computer science in China and Canada, I joined this project with an epistemological stance shaped by positivism. I had been trained in the positivist way without realizing it, until this project. As the project unfolded, tensions between my training and the realities of community-based research prompted me to be aware of and interrogate my epistemological assumptions, leading to an evolution in my epistemological perspective. As the first author, many of the reflections presented in this paper are informed by my evolving journey.

\textbf{Foroozan Daneshzand}:
I am a postdoctoral researcher in HCI and information visualization with a background in design, and much of my work focuses on participatory and community-engaged approaches to data and visualization. I entered this project interested in how visualization could support communities in understanding and communicating data that mattered to them. Through working closely with the arts community, I became increasingly aware that what counts as relevant data, how problems are framed, and what constitutes a useful visualization cannot be separated from the perspectives and lived experiences of the people involved. This experience made me more attentive to how our understanding of the problem, the data, and possible design directions continuously shape one another throughout co-design.

\textbf{Zezhong Wang}:
I am a postdoctoral researcher in information visualization and HCI, with a background in user experience design and data-driven storytelling. My work is motivated by making data more understandable and engaging through visual narrative and interaction, and I entered this project inclined to see design and technical prototyping as ways to translate complex data into accessible forms. Working with Vancouver Island arts communities unsettled this assumption: I came to understand that accessibility cannot be separated from who defines and collects the data, whose experiences are absent, and how visualization redistributes agency. As a visualization researcher and designer, I held disciplinary and technical authority within the collaboration; this project has made me more attentive to listening, remaining accountable, and creating space for community partners to question and reshape the data and visualizations, rather than positioning them merely as recipients of a finished system.

\textbf{Erica Mattson}: I am an artist, researcher, and strategist whose work sits at the intersection of systems change and collaborative practice. My perspective has been shaped by more than two decades of working alongside artists, communities, and public institutions, as well as by graduate studies in applied communication and Nora Bateson’s Warm Data practice. Being involved in this work has helped me articulate something I had long sensed through practice: that every dataset reflects choices about what becomes visible and what remains absent, and that collective sense-making emerges through the relationships people develop with one another and with the data. For me, this project did not introduce entanglement as a theory; it gave language to patterns I had been navigating throughout my professional and artistic practice. I hope this work contributes, in some small way, to the much longer journey of building more equitable and sustainable systems.

\textbf{Jenny Farkas}:
I am a creative economic developer, social entrepreneur, community animator, co-founder of Vancouver Island’s Creative Coast collaboration lab, and founder of MakeSpace for Art Society in Victoria, British Columbia. My perspective in this research is grounded in more than two decades of work in communications, place branding, community development, and, increasingly, the creative economy across Vancouver Island. I came to this collaboration as a practitioner and community partner rather than an academic researcher, bringing relationships, lived knowledge of the local creative sector, and a strong interest in how data can serve communities rather than simply describe them. My work is informed by a commitment to economic inclusion and a belief that talent is widely distributed while access to the conditions that allow it to flourish is not. These experiences shaped both the questions I brought to the project and my understanding of what constitutes useful knowledge, meaningful participation, and successful community-based research.

\textbf{Sarah Storteboom}:
I am a trans-disciplinary data storyteller, and was involved in the early stages of this project. I have worked in the data visualization field for eight years and have a background in visual arts. My art practice, which centered around participatory public art, led me to utilizing technology for data processing and visualization.That is how the two worlds connected for me. I am deeply influenced by my artistic roots in relation to data. I embrace the human elements of the data pipeline and think of data as something to be expressed. I have a multi-lensed epistemological view and it was delightful to work with artists in this project who were seeing data through the lenses of their own practices and desires.

\textbf{Charles Perin}:
I am an Associate Professor of Computer Science at the University of Victoria, BC, Canada. I have 15 years of training and experience in human-computer interaction and data visualization. Until 10 years ago, my thinking was limited to positivism; Then I progressively learned about, practiced, and valued other ways of knowing. Yet, this project brought to me unique insights and reflections as the boundaries of research and practice, researcher and participant, author and `stakeholder', collaborator and `domain expert', slowly fell apart. What started with a `design study' mindset turned out to be a profound reflection on the way I conduct visualization research, its framings, its words, its dark sides and its entrenched epistemologies.

\textbf{Sheelagh Carpendale}:
I am a Professor and Canada Research Chair in Data Visualization at Simon Fraser University. While I now have over 30 years of experience in data visualization and human-computer interaction, I started from Art School and Design School some 50 years ago. Within the international VIS research community, I have operated somewhat closeted from an interpretivist and critical perspective. This project has led me to deeply question my willingness to let positivist VIS norms dominate such things as my vocabulary and practices. The experience of this project has led me to see the importance of not only recognizing but following through with more than the acceptance of alternate ways of knowing, to the appreciation, the surfacing and the celebration of all the different types of knowledges and understandings. In fact, I am trying to embrace J. Grahn's suggestion that rather than \textit{under}standing, we aim for \textit{inter}standing or better yet \textit{intra}standing~\cite{grahn1989really}. 



\section{Conclusion} 


By examining and reflecting on a longitudinal community-based visualization co-design project through the lens of entanglement theory, we showed how this alternative epistemological perspective enabled us to reframe epistemic tensions from research challenges into opportunities for methodological and theoretical insight. In doing so, we critically examined the language of visualization research, developed a more nuanced understanding of visualization co-design, and exposed limitations of conventional visualization design pipelines.
We hope this work encourages visualization researchers to engage more explicitly with the epistemological assumptions underlying their practices and to explore alternative epistemologies as both critical and generative lenses.

\acknowledgments{
This research was funded in part by NFRFR-2022-00570 (A Co-Design Exploration), NSERC Discovery Grant: RGPIN-2019-07192 and RGPIN-2019-05422, and Canada Research Chair in Data Visualization CRC-2019-00368. }
\bibliographystyle{abbrv-doi-hyperref}

\bibliography{template}

\end{document}